\pdfoutput=1

\documentclass[aps, prb, twocolumn, showpacs,floatfix, groupedaddress]{revtex4}
\usepackage{bm}
\usepackage{graphicx}
\usepackage{pdfpages}

\usepackage[utf8]{inputenc}
\usepackage[english]{babel}
\usepackage{natbib}
\begin{document}

\title{High resolution Kerr microscopy study of exchange bias phenomena in FePt/Fe exchange spring magnets}

\author{Zaineb Hussain}
\affiliation{UGC-DAE Consortium for Scientific Research, University Campus, Khandwa Road, Indore 452001, India.}

\author{Dileep Kumar} 
\affiliation{UGC-DAE Consortium for Scientific Research, University Campus, Khandwa Road, Indore 452001, India.}

\author{V. Raghavendra Reddy} 
\email{varimalla@yahoo.com; vrreddy@csr.res.in}
\affiliation{UGC-DAE Consortium for Scientific Research, University Campus, Khandwa Road, Indore 452001, India.}

\author{Ajay Gupta} 
\affiliation{Amity Center for Spintronic Materials, Amity University, Noida 201303, India.}

\begin{abstract}

Magnetization and magnetic microstructure of top soft magnetic layer (Fe), which is exchange spring coupled to bottom hard magnetic layer ($L1_0$ FePt) is studied using high resolution Kerr microscopy. When the sample (FePt/Fe) is at remanent condition of hard magnetic layer, considerable shifting of Fe layer hysteresis loop from centre i.e., exchange bias phenomena is observed. It is observed that one can tune the magnitude of exchange bias shift by reaching the remanent state from different saturating fields ($H_{SAT}$) and also by varying the angle between measuring field and $H_{SAT}$. The M-H loops and domain images of top soft Fe layer demonstrates unambiguously that soft magnetic layer at remanent state in such exchange coupled system is having unidirectional anisotropy.  An analogy is drawn and the observations are explained in terms of the mostly accepted models of exchange bias phenomena exhibited by bilayers consisting of ferromagnetic(FM) and anti-ferromagnetic (AFM) layers, when the AFM layer is field cooled across $N\acute{e}el$ transition temperature. 
 
\end{abstract}

\keywords{Exchange spring magnets, exchange bias, magnetic domains, Kerr microscopy}

\pacs{75.60.-d, 75.60.Ch}

\maketitle

\maketitle \section{Introduction}
Exchange bias (EB) phenomena, which manifests as a shift of hysteresis loop along the field axis, occurs due to anisotropy induced across the ferromagnetic (FM) – antiferromagnetic (AFM) interfaces when the system is field-cooled through the $N\acute{e}el$ temperature ($T_N$) of the AFM \cite{JMMMReview}. Several applications of these materials such as permanent magnets, recording media, spin-valves etc., are demonstrated and are being utilized \cite{JMMMReview}. Since EB phenomena involves a complicated interfacial magnetic interaction between FM and AFM layers, one can expect several spin geometries depending on bilayer materials, exchange coupling strength of the bilayers etc., manifested as different magnetization reversal process \cite{JMMMReview}. In most of the cases, due to EB phenomena,  asymmetric shaped hysteresis loops are observed and methods such as Lorentz microscopy, Kerr microscopy etc., are used to study the magnetic domain evolution in such FM-AFM exchange biased systems \cite{Lorentz, RudiJAP}. The degree of loop or domain asymmetry found in these experiments is very different, in some cases even it is almost negligible \cite{Qian}. Typically, for the asymmetric loops, two magnetization reversal mechanisms are  observed viz., coherent rotation, nucleation and domain wall propagation \cite{RudiJAP}. 

Field cooling protocol from above the magnetic transition of AFM layer ($T_N$) sets a pre-biasing of the interface moments, which essentially result in shift along the field axis generally in the opposite direction to the cooling field. This shift i.e., EB phenomena decreases as a function of temperature and repeated application of magnetic field (known as training effect) at a given temperature.  The temperature at which shift becomes zero i.e., the point at which EB phenomena disappears is usually called the blocking temperature ($T_B$). For example, in Co-CoO based compounds which are extensively explored in literature, $T_N$ is 290 K and $T_B$ is found to be as low as 200 K and therefore limiting their room temperature (RT) based application capability \cite{JMMMReview}. Whereas $T_B$ values for metallic AFM layers such as $Fe_{50}Mn_{50}$ are found to be above RT and are more attractive \cite{JMMMReview}.  Therefore, from an application point of view, the system should exhibit EB phenomena at and above RT.   

Recent studies indicate that an EB phenomena is not only restricted to FM-AFM, but can also be observed in exchange-coupled magnetic layers with different magnitudes of coercive fields such as exchange spring magnets \cite{Klein}. The advantage of exchange spring magnets as compared to FM-AFM systems is that one need not to field cool the samples across $T_N$ for obtaining EB phenomena and are functional even at room temperatures.  Apart from this, exchange spring magnets consisting of soft magnetic (SM) and hard magnetic (HM) phases are  the focus of current research because of various technological applications such as permanent magnets with high energy products, ultra-high density recording etc \cite{Coey, Victoria}. Usually in such coupled layers, the response of the SM layer and also the evolution of magnetic microstructure are different as compared to individual and un-biased SM layers.  Kerr microscopy is one technique which can give both magnetization and magnetic microstructure in a single measurement and therefore is being used extensively in recent times \cite{RudiBook, McCordReview}.  In the present work, we report the Kerr microscopy investigation of exchange spring coupled FePt/Fe bilayer prepared by ion beam sputtering.

\maketitle \section{Experimental}
FePt/Fe bilayer film with the structure FePt(30nm)/Fe(22nm)/C(2nm) was deposited by ion beam-sputtering and detailed structural, magnetic and exchange spring behavior are  reported in ref \cite{VRRTSF}.  The FePt layer was in $L1_0$ phase i.e., exhibiting hard magnetic properties \cite{VRRTSF}. The magnetic domains were imaged by high resolution magneto-optical Kerr microscope (M/s Evico Magnetics, Germany) and the magnetization (M-H) loops were measured simultaneously by deriving the magnetization signal from the average domain image intensity \cite{RudiBook, McCordReview}. The measurements are carried out in longitudinal and transverse mode.  For the sake of comparison, an Fe film of about 22 nm thickness is deposited on silicon substrate and is also studied with Kerr microscopy.

\begin{figure}[htbp]
  \centering
  \includegraphics[width=\linewidth]{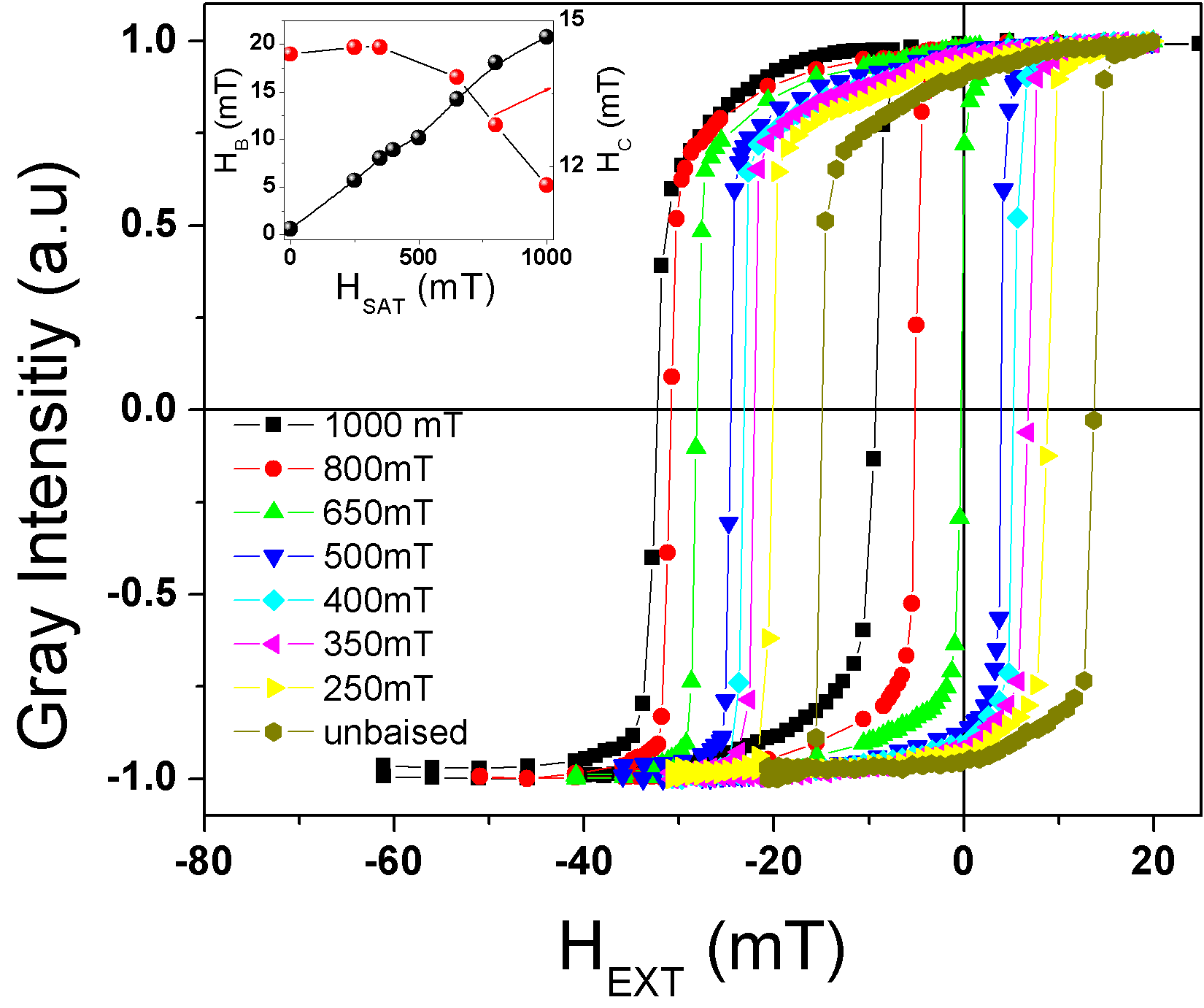}
  \caption{(a) Longitudinal MOKE hysteresis loops of top Fe layer in FePt/Fe sample measured at remanent state approached from different saturating fields ($H_{SAT}$).  Note the horizontal shift ($H_B$) of loops. Unbiased correspond to the case when remanence approached with $H_{SAT}$ made to zero in oscillating mode. (b) Variation of shift ($H_B$) and coercivity ($H_C$) with $H_{SAT}$. Solid line is guide to eye.}
\end{figure}

\begin{figure}[htbp]
  \centering
  \includegraphics[width=\linewidth]{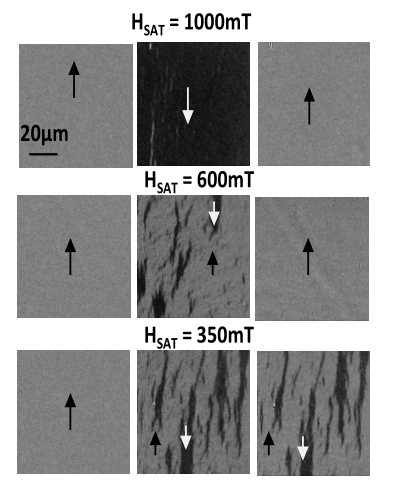}
  \caption{Kerr domain images of top Fe layer in FePt/Fe sample captured  along the M-H loop of Figure-1 for different indicated $H_{SAT}$. (left) at saturating field i.e., background image, (middle) near the switching field and (right) at remanence condition. The arrows  show the direction of magnetization. Scale bar is same for all the images.}
\end{figure}

\maketitle \section{Results and Discussions}

Figure-1(a) shows the M-H loops of the top soft Fe layer recorded at remanent condition of the sample.  The remanent state of the sample is reached by subjecting the sample to different applied field strengths i.e., the FePt/Fe bilayer was initially subjected to say 1000 mT field (saturation field used for reaching remanent state, henceforth designated as $H_{SAT}$) and then slowly $H_{SAT}$ was made to zero. For the subsequent measurement the sample was subjected to an oscillating zero field. As one can see from Figure-1, the loops are shifted to negative side when the remanent state was approached from positive $H_{SAT}$.  Reverse behavior i.e., shifting of loop to positive side is observed when the remanent state was approached from negative $H_{SAT}$. This is similar to the case of EB effect observed in FM-AFM systems \cite{JMMMReview}. In other words instead of having two easy equivalent configurations in opposite directions the Fe layer is having a preferred easy axis parallel to the bottom hard magnetic FePt layer magnetization at the remanent state i.e., exhibiting uni-directional anisotropy, which is proved in the last section (Figure-7). The magnitude of shifting ($H_B$) is found to depend significantly as shown in Figure-1(b) on the strength of $H_{SAT}$. This can be explained by realizing that the bottom hard magnetic FePt layer which is coupled to top Fe layer, is still aligned parallel to $H_{SAT}$ at remanence and therefore one need to apply more fields in opposite direction as that of $H_{SAT}$ to switch the Fe magnetization.
  
	\begin{figure}[htbp]
  \centering
	\includegraphics[width=\linewidth]{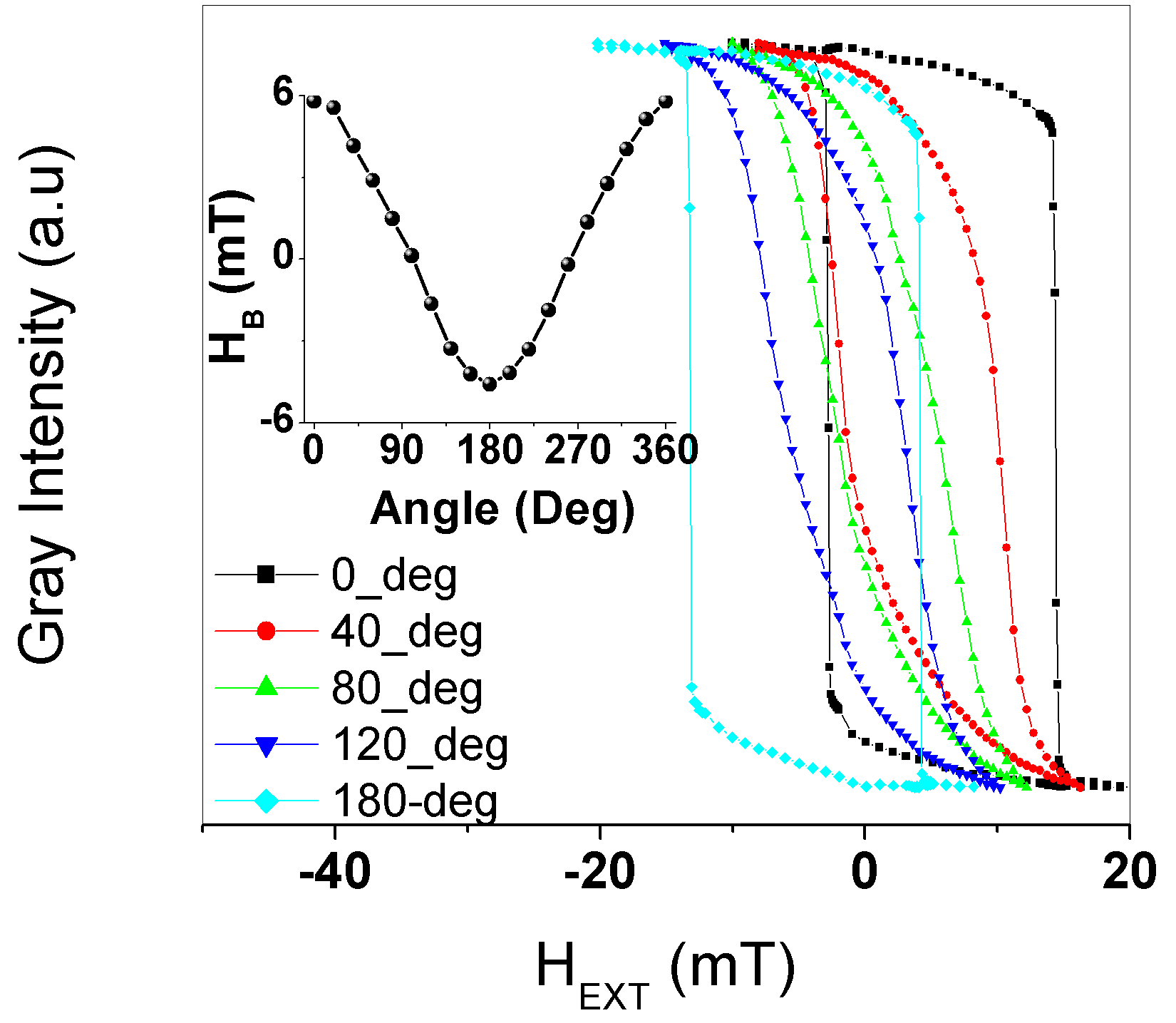}
  \caption{Longitudinal MOKE hysteresis loops of top Fe layer in FePt/Fe sample measured at remanent state approached from -330 mT ($H_{SAT}$) and inset show the plot of variation of biasing field ($H_B$) with angle between $H_{SAT}$ and applied measuring field ($H_{EXT}$).}
\end{figure}

\begin{figure}[b]
  \centering
  \includegraphics[scale=1, width=\linewidth]{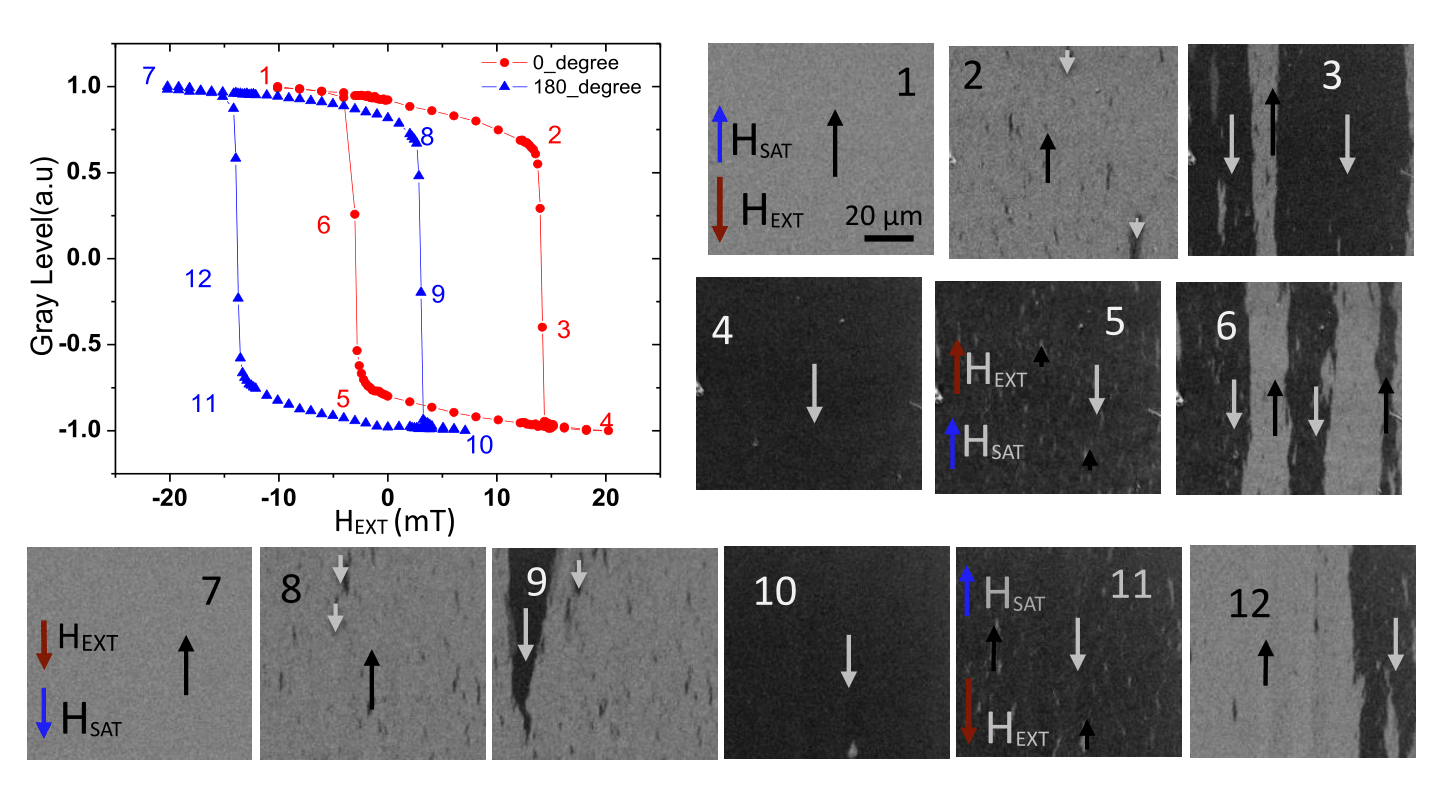}
  \caption{Longitudinal MOKE hysteresis loops of top Fe layer in FePt/Fe sample measured along and at $180^o$ to the $H_{SAT}$ with the corresponding domain images captured along the M-H loop at the indicated points. Scale bar is same for all the images.}
\end{figure}	
		
To give insight, the Kerr images are recorded along the M-H loop of Figure-1. Initially the sample is saturated (here the saturation field is for the purpose of recording the minor loop of top Fe layer i.e., about -60 mT is applied as shown in Figure-1) and the background image is captured (shown in the left frame of Figure-2). Then the external field is increased in the opposite direction till the field is slightly above the coercive field. The domains are found to nucleate indicating the switching of Fe layer magnetization (the corresponding image is captured and is shown in middle frames of Figure-2).  Then the external field is made zero and the domain images are captured at this state which is called as remanent images henceforth (as shown in right frames of Figure-2).  One can clearly see from the images that the remanent images corresponding to higher $H_{SAT}$ are same as that of background images i.e., Fe layer magnetization is going back to the initial starting point which is nothing but the spring effect.  This clearly indicates that the top Fe layer is more effectively pinned / coupled with bottom FePt layer for the case of higher $H_{SAT}$ and therefore one needs to apply more external field to switch Fe magnetization, explaining the increasing value of $H_B$ with increasing $H_{SAT}$.   

\begin{figure}[htbp]
  \centering
  \includegraphics[scale=1, width=\linewidth]{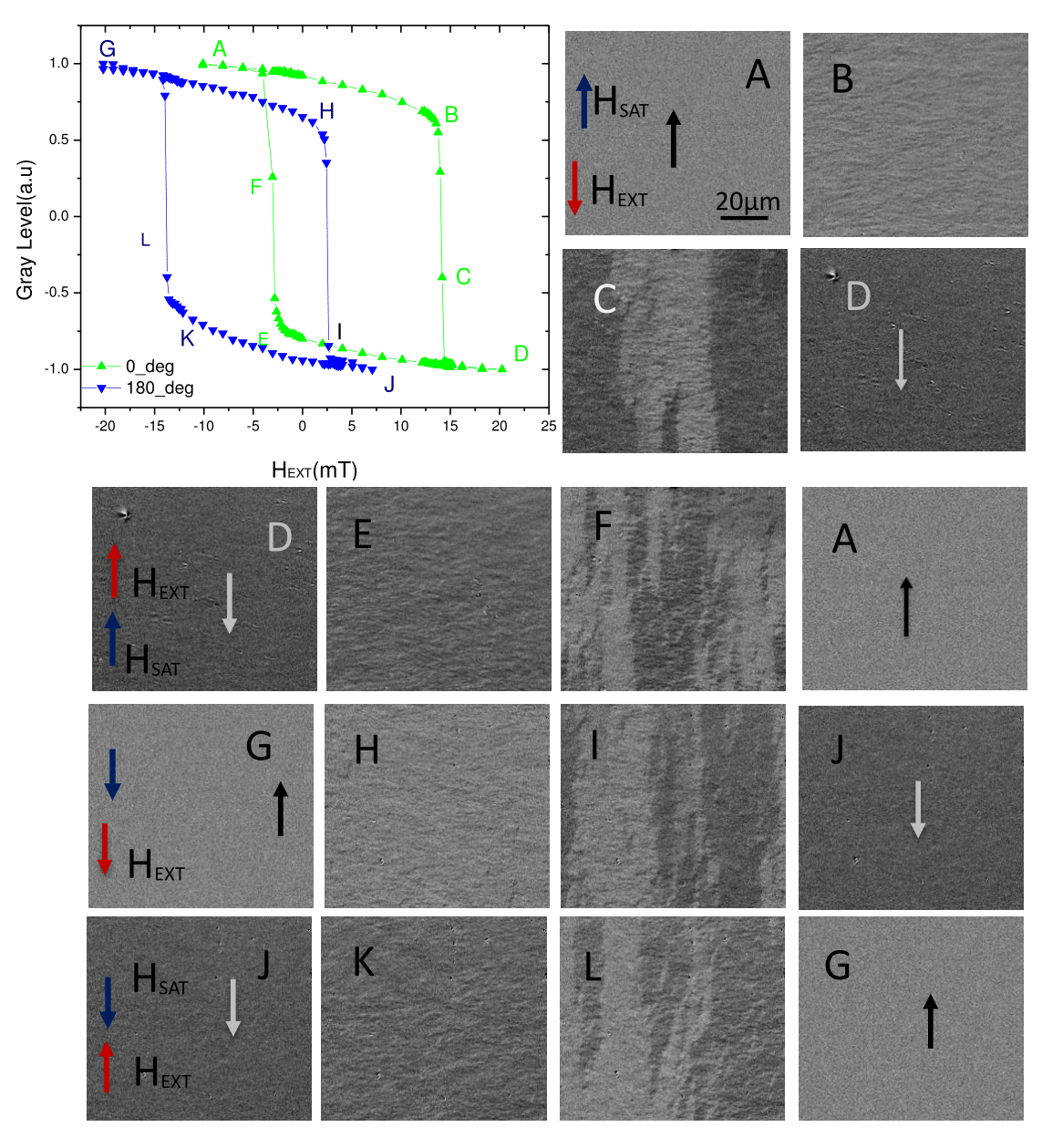}
  \caption {Transverse MOKE hysteresis loops of top Fe layer in FePt/Fe sample measured at $0^o$ and $180^o$ to the $H_{SAT}$ with the corresponding domain images captured along the M-H loop at the indicated points. The arrows  show the direction of magnetization. Scale bar is same for all the images.} 
\end{figure}

Another relevant observation is that the coercivity ($H_C$) is found to depend on the strength of the $H_{SAT}$ as shown also in Figure-1(b).  Initially the $H_C$ is found to be almost constant but for fields of $H_{SAT}$ $\geq$ 350 mT, $H_C$ is found to decrease accompanied by increase of $H_B$ with increasing $H_{SAT}$.  These results can be understood in terms of well accepted models of coupled FM-AFM layers exhibiting EB phenomena.  For example, in spin-glass model, the magnetic state of the interface between the FM layer and AFM layer is assumed to be magnetically disordered similar to that of a spin glass system. According to the spin-glass model, AFM layer is assumed to consist of two states viz., one part which has larger anisotropy and the second part has a weaker anisotropy \cite{ThesisRadu}. Intermixing, stoichiometry deviations, roughness etc., at the interfaces are considered to be responsible for the formation of such low anisotropy region. Fraction of such low anisotropy interfacial spins can rotate almost in phase with that of FM spins and therefore are frustrated leading to enhanced coercivity.  On the other hand, the more the high anisotropy AFM fraction, the more will be the shift of loop i.e., higher values of $H_B$ and less would be the coercivity \cite{ThesisRadu}.  Extending this formalism to the present case, when the bottom HM FePt layer is aligned by applying a field of 1 Tesla, all the FePt spins are aligned in the applied field direction and hence can be considered that the interface disorder is low.  It is to be noted that a field of about 1 Tesla is expected to saturate the studied sample as evidenced from bulk magnetization measurement \cite{VRRTSF}.   Whereas, if the applied field ($H_{SAT}$ $<$ 1 Tesla) is not enough to align all the FePt spins, the interface is expected to consist of some aligned and random FePt spins i.e., in a disordered state leading to the increased $H_C$ and decreased $H_B$ as observed.

Further in order to study the magnetic microstructure Kerr microscopy measurements are carried out on the top soft Fe layer. The FePt/Fe bilayer was kept in remanent condition from $H_{SAT}$ equivalent to -330 mT. This value of field is selected due to instrumental constraints so as to accommodate the high resolution objective lens and to perform angular dependent measurements. The sample was rotated in-plane i.e., the measurements were carried out as a function of angle between the applied external magnetic ($H_{EXT}$) field used for recording the M-H loop and $H_{SAT}$ as shown in Figure-3. The observed azimuthal variation of $H_B$, as shown in the inset of Figure-3,  matches with that of FM-AFM systems exhibiting exchange bias phenomena \cite{ThesisRadu, JPCM2006, LSun2005}. Further, it is to be noted that no decrease in $H_B$ is observed when the measurements are carried out repeatedly i.e., no training effect is observed in the present work. Even when the $H_{SAT}$ was equal to 1 Tesla no training effect was observed.  It is reported in FM-AFM systems exhibiting EB phenomena that, if the AFM is highly anisotropic then the training effect could be absent \cite{ThesisRadu}. Bottom $L1_0$ FePt layer is known to possess high magnetic anisotropy \cite{VRRTSF} and therefore could be the reason for the absence of training effect.  However, some more experiments such as FePt layer with different degree of chemical ordering and hence different anisotropy, different thickness of SM and HM layers etc., are being planned to comment conclusively on this aspect. 

\begin{figure}[t]
  \centering
	\includegraphics[scale=1, width=\linewidth]{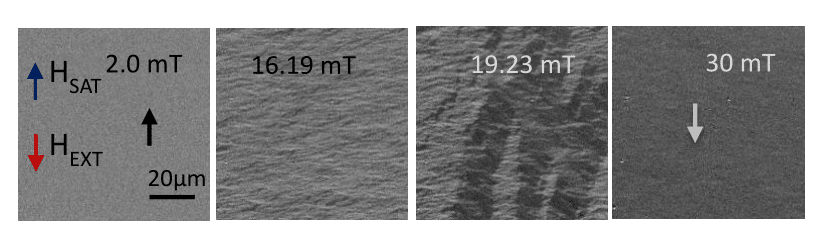}
  \caption {Transverse MOKE domain images of top Fe layer in FePt/Fe sample measured when $H_{SAT}$=1 Tesla.  The arrows  show the direction of magnetization. Scale bar is same for all the images. The images are captured during M-H loop measurement similar to that of Figure-3(A)-(D) for the case of $H_{SAT}$=1 Tesla.} 
\end{figure}

\begin{figure}[htbp]
  \centering
  \includegraphics[scale=1, width=\linewidth]{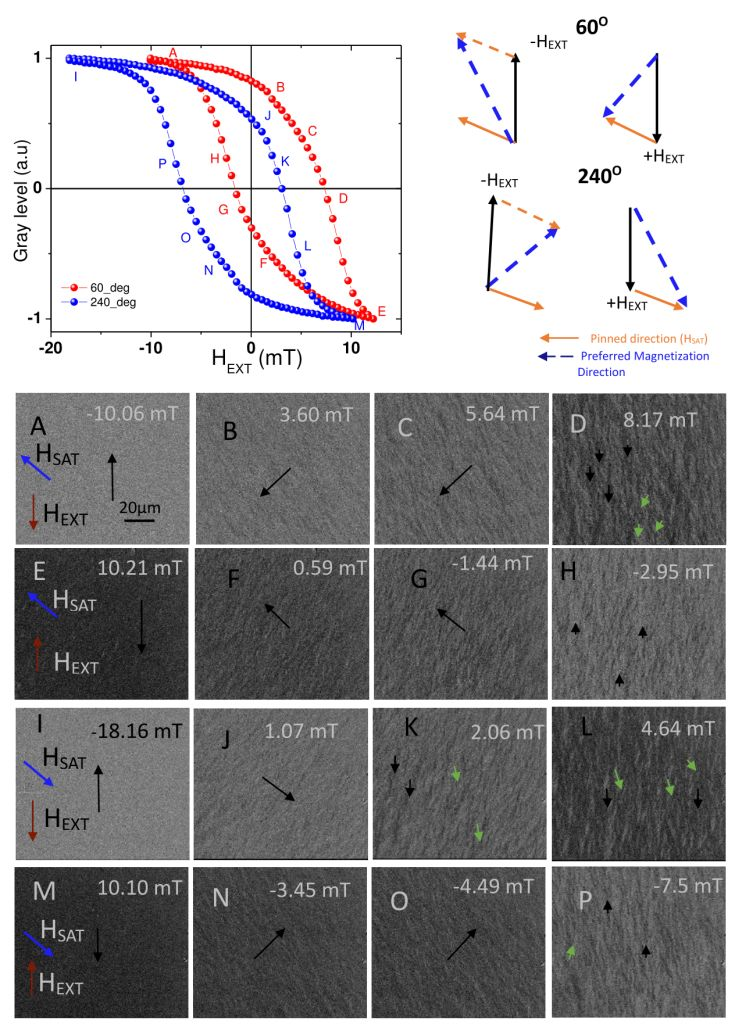}
  \caption{Longitudinal MOKE hysteresis loop and domain images captured along the M-H loop of top Fe layer in FePt/Fe sample measured at $60^o$ and $240^o$  with respect to $H_{SAT}$. Scale bar is same for all the images.  The arrows  show the direction of magnetization. Top right show the pictorial diagram showing the competition between the two forces acting viz., $H_{EXT}$ and $H_{SAT}$ to explain the domain evolution.} 
\end{figure}

At each angle between $H_{EXT}$ and $H_{SAT}$, magnetic domains are recorded along with M-H loops as discussed following. Figure-4 shows the M-H data and magnetic microstructure i.e., domain images captured at different positions of M-H loop for the angle of $0^o$ and $180^o$ i.e., the $H_{EXT}$ is parallel and anti-parallel to $H_{SAT}$. One can clearly see that the magnetization reversal is by domain nucleation and growth mechanism for both forward and backward branches of magnetization loop. However, when the Kerr microscopy measurements are carried out in transverse geometry, unlike the longitudinal case, the observed domain patterns exhibit ripple formation  along both the forward and backward branches as shown in Figure-5, with the modulation being orthogonal to average magnetization \cite{Smith1962, RippleTEM}. This is same even for the case of $H_{SAT}$=1 Tesla as shown in Figure-6. The observed ripple like feature could be indicative of magnetization twisting of soft magnetic layer in exchange spring magnet systems as reported by Chumakov et al., and McCord et al \cite{Chumakov, McCord}.  Using Kerr microscopy, McCord et al., have reported chirality reversal for planar interface domain walls in exchange-coupled hard/soft magnetic $Tb_{45}Fe_{55}$(50 nm) /$Gd_{40}Fe_{60}$(50 nm) bilayers combined with magneto-resistance measurements and modelling \cite{McCord}.  It is reported that domain nucleation and lateral domain wall propagation play key role in the chirality reversal.  Similarly, Chumakov et al., \cite{Chumakov} have studied the magnetization process in $Sm_{40}Fe_{60}$ (88 nm)/$Ni_{80}Fe_{20}$ (62 nm) exchange spring films.  Formation of $180^o$ domain state is reported along the easy axis at remanent state. Textured contrast i.e., magnetization ripple orthogonal to the field direction is observed along the reversible branch of hysteresis loop followed by the nucleation and growth of regular domains along irreversible branch \cite{Chumakov}.  In the present work also, one can see that initially with increasing field the observation of ripple like features indicative of  magnetization twisting of soft magnetic layer as reported by Chumakov et al., followed by domain nucleation and growth. With increase in $H_{EXT}$ the domains starts to nucleate and grow whose magnetization direction is parallel to external field direction as shown in Figure-5 (C), (F), (I) and (L).  It is observed that modulation pattern (not shown here) remains even after the domain nucleation and growth.  Ripples disappear at sufficiently higher fields, as high as close to saturation of the loops.  

\begin{figure}[b]
  \centering
  \includegraphics[scale=1, width=\linewidth]{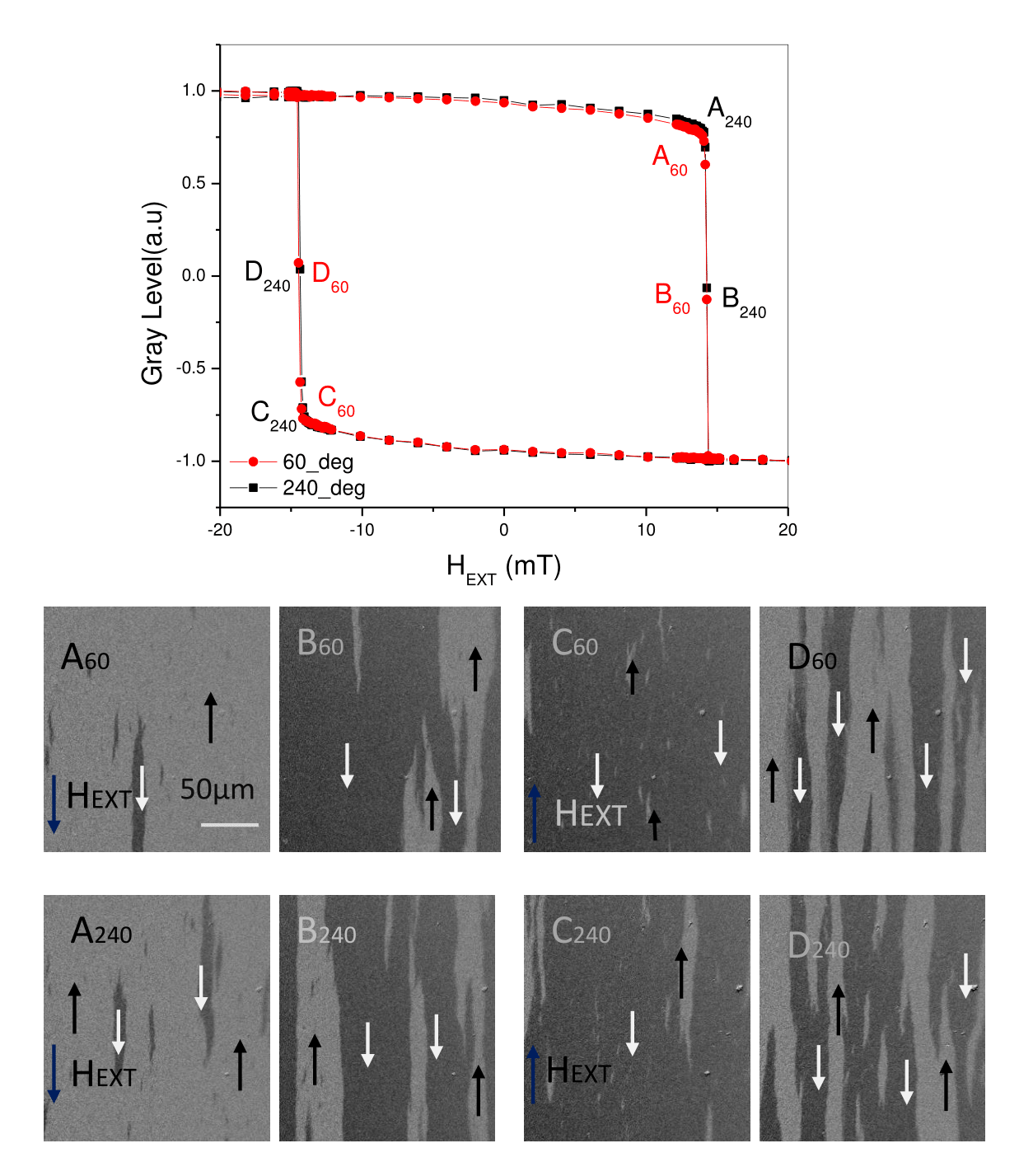}
  \caption{Longitudinal MOKE hysteresis loop and domain images captured along the M-H loop of Fe film measured at $60^o$ and $240^o$. Scale bar is same for all the images. The arrows show the direction of magnetization.} 
\end{figure}

Figure-7 shows the M-H data and magnetic microstructure for the angle of $60^o$ and $240^o$ between $H_{EXT}$ and $H_{SAT}$.  One can see from the data that the magnetization reversal for both forward and backward branches have considerable changes both in domain structure and hysteresis loop.  Reversal is taking place by magnetization rotation by forming ripple pattern and the modulations are found to be reproducible.  From Figure-7(A) to (C) images  magnetization moments are aligned in certain direction with respect to $H_{EXT}$. Gradually it  rotates with the application of $H_{EXT}$ and then as shown in image Figure-7(D), the domains whose magnetization direction is in  $H_{EXT}$ direction starts growing and then gradually these ripple pattern are wiped by these domains. Same pattern is observed in images Figure-7(E) to (H) which corresponds to backward branch of magnetization loop. Therefore, for both the branches the domain evolution is same i.e, first by magnetization rotation and then by domain wall motion.  The same is the case for the angle of $240^o$.  But the difference for both forward and backward branches at a given angle, lies in  magnetization ripple direction i.e, both are opposite corresponding to initial points (Figure-7(B), (F); Figure-7(J), (N)). This can be understood by considering the competition between the two forces acting viz., $H_{EXT}$ field which is trying to pull along its direction and other one is  the pinning direction or biasing direction $H_{SAT}$ which is now $60^o$ and $240^o$ as shown in Figure-7.  The resultant direction of  magnetization will be governed by this competition merely. This gives us a clear evidence that the top SM Fe layer is pinned with bottom HM FePt layer and hence explains the observed EB phenomena.   Otherwise for both forward and backward branches the magnetization ripple direction should have been same.  This statement is substantiated by recording the Kerr microscopy data of free and un-biased Fe layer of about 22 nm thickness. The observations are typical for a soft magnetic film exhibiting uniaxial anisotropy. For all the in-plane angles, the magnetization reversal is by domain nucleation and growth. Figure-8 show the M-H data along with domain images measured at $60^o$ and $240^o$ deg (the angle between the easy axis and the applied field for measuring M-H loop i.e., $H_{EXT}$).  One can clearly see the difference in domain evolution of free and biased Fe layer by comparing the data of Figure-7 and Figure-8.  Therefore, the modulation pattern observed for the case of FePt/Fe layers (Figure-7) is because of uni-directional anisotropy of the SM Fe layer as a consequence of the competition of $H_{SAT}$ and $H_{EXT}$. 

\maketitle \section{Conclusions}

In conclusion, Kerr microscopy study of top soft magnetic Fe layer is reported in exchange spring coupled $L1_0$ FePt(30 nm)/Fe(22nm) bilayer. The results demonstrate that one can have exchange bias phenomena i.e., shifting of hysteresis loops in one direction at remanent state of such exchange spring systems.  Also, one can tune the magnitude of exchange bias shift by reaching the remanent state from different saturating fields ($H_{SAT}$) and also by varying the angle between measuring field and $H_{SAT}$. The presented M-H loops and domain images of top soft Fe layer conclusively confirms that soft magnetic layer at remanent state in such exchange coupled system is having unidirectional anisotropy. Two phenomena i.e., exchange bias and exchange spring effect exhibited by the same structure at room temperature might pave way for variety of applications. 

\maketitle\section {Acknowledgments}
VRR thank Prof. Rudolf Schafer for the discussions and reading the manuscript. Thanks are due to the Director and Centre Director of UGC-DAE CSR, Indore for the encouragement. 

\bibliography{manuscript} 

\end{document}